\documentclass[aps,amsmath,amssymb,amsfonts,superscriptaddress,floatfix]{revtex4} 
\pdfoutput=1
\usepackage{amsmath,amsthm,amsfonts,amssymb}
\usepackage{graphicx}

\usepackage{color}
\numberwithin{equation}{section}
\numberwithin{figure}{section}

\DeclareMathAlphabet{\mathpzc}{OT1}{pzc}{m}{it}

\newcommand{\be}{\begin{equation}}
\newcommand{\ee}{\end{equation}}

\newcommand{\nlog}{K}

\newcommand{\nlm}{\nlog_{max}}
\newcommand{\nphys}{N_{maj}}
\newcommand{\nstab}{N_{stab}}
\newcommand{\nqub}{N_{qub}}
\newcommand{\li}{m}
\begin{document}

\title{Small Majorana Fermion Codes}

\author{Matthew B.~Hastings}

\affiliation{Station Q, Microsoft Research, Santa Barbara, CA 93106-6105, USA}
\affiliation{Quantum Architectures and Computation Group, Microsoft Research, Redmond, WA 98052, USA}

\begin{abstract}
We consider Majorana fermion stabilizer codes with small number of modes and distance.  We give an upper bound on the number of logical qubits for distance $4$ codes, and we construct Majorana fermion codes similar to the classical Hamming code that saturate this bound.  We perform numerical studies and find other distance $4$ and $6$ codes that we conjecture have the largest possible number of logical qubits for the given number of physical Majorana modes.  Some of these codes have more logical qubits than any Majorana fermion code derived from a qubit stabilizer code.
\end{abstract}
\maketitle
Qubit stabilizer codes are a fundamental way to construct families of quantum error correcting codes.  These codes
use some number, $\nqub$, of physical qubits, to construct some smaller number, $\nlog$, of logical qubits.
The code space is the $+1$ eigenspace of several mutually commuting operators.  These operators are called stabilizers, and are taken to be products of Pauli operators\cite{stab}.
Majorana fermion codes, introduced in Ref.~\onlinecite{blt}, are a natural variant of stabilizer codes where the
stabilizers are instead taken to be products of Majorana operators.  Instead of using qubits as the physical degrees of freedom, the Majorana codes use some number, $\nphys$, of Majorana modes, to obtain a code space with $\nlog$ logical qubits (see below for identification of the code space with qubits).

In Refs.~\onlinecite{anyons,blt}, it was shown how to convert qubit stabilizer codes into Majorana fermion codes, with the properties of the Majorana fermion codes (including distance, number of logical qubits, and weight of generators) depending on those of the original stabilizer code.  
Further, Ref.~\onlinecite{blt} discussed various other Majorana fermion codes which could not be obtained by such a conversion procedure.

In this paper, we further consider Majorana fermion codes which cannot be obtained from a qubit stabilizer code.
However, our focus will be on {\it small} codes.  That is, rather than studying asymptotic properties with large number s of physical Majorana modes, we will instead consider codes that have small $\nphys$ and obtain optimal distance  $d$ for the given $\nlog$.  To explain by analogy to qubit stabilizer codes, our study will be closer to the results in the code tables of Ref.~\onlinecite{codetables}, rather than studying topological phases such as toric codes or color codes.

One motivation for studying small Majorana fermion codes is that hopefully realizations of Majoranas in physical devices\cite{karzig} will have very low error rates.  Perhaps these modes will already have low enough error rates that no code will be necessary, but if a code is necessary, then a low distance code may suffice.

In some cases, we will be able to prove that our small Majorana fermion codes have an optimal tradeoff between $\nphys,d,\nlog$.  These codes will be closely related to Hamming codes.
In other cases, we will conduct computer search to construct codes that we conjecture have an optimal tradeoff; the computer search will not be exhaustive but will involve a random element, so we will not be able to prove optimality.
We consider only the case where the codes have no odd weight logical operators, as explained below.

\section{Majorana Stabilizer Codes}
\subsection{Hilbert Space, Code Space, and Stabilizer Group}
We consider a system with $\nphys$ Majorana fermion operators.  We denote these Majorana fermion operators by $\gamma_a$ with $a\in {1,\ldots,\nphys}$.
They obey the anti-commutation relations
\be
\{\gamma_a,\gamma_b\}=2 \delta_{a,b}.
\ee
We will always assume that $\nphys$ is even.
The minimal Hilbert space compatible with these anti-commutation relations has dimension $2^{\nphys/2}$ and we will take this to be the dimension of the Hilbert space of the system.
A Majorana fermion code is a subspace of this Hilbert space.

We will consider Majorana fermion codes which have a stabilizer form, so that there are several operators, called ``stabilizers", such that the code space (the subspace of the Hilbert space which describes valid codewords) is the space in which each of these operators assumes some given eigenvalue.  Each of these operators will be a product of an {\it even} number of Majorana fermion operators; physically, this is chosen so that they correspond to bosonic operators.  If the number of operators in the product is equal to $2 \mod 4$, then the operator is anti-Hermitian and the possible eigenvalue are either $+i$ or $-i$ while if the number of operators in the product is equal to $0 \mod 4$ then the operator is Hermitian and the possible eigenvalues
are either $+1$ or $-1$.
Further, all of these operators will be chosen to commute with each other.
Thus, as an example code, one might take a system with $\nphys=6$ and with stabilizers $\gamma_1 \gamma_2 \gamma_3 \gamma_4 \gamma_5 \gamma_6$ and $\gamma_1 \gamma_2$ (this code is practically useless as it has distance $2$ as defined below, but it is a valid code).

The stabilizers generate a group, the stabilizer group, which is the group generated by products of stabilizers.  Taking the quotient of this group by
all elements of the group which are proportional to the identity (i.e., all elements equal to $1,-1$) gives a group with $2^{\nstab}$ elements,
where $\nstab$ is the minimal number of stabilizers that generate this group.  That is, if for example one were given a list of stabilizers $\gamma_1 \gamma_2 \gamma_3 \gamma_4 \gamma_5 \gamma_6$ and $\gamma_1 \gamma_2$ and $\gamma_3 \gamma_4 \gamma_5 \gamma_6$ then the group has $\nstab=2$ (despite the fact that there were $3$ stabilizers in the list) as it is generated by $2$ stabilizers (indeed, any two stabilizers from that list suffices).

One way to understand this group is to identify each element of the stabilizer group with a bit string of length $\nphys$.  There will be a $1$ in the $a$-th entry of the bit string if the operator $\gamma_a$ is in the given element of the stabilizer group.  Thus, with $\nphys=6$, the operator $\gamma_1 \gamma_2$ will correspond to the string $110000$.  The sign of the operator is irrelevant to the bit string, so that $\gamma_1\gamma_2$ and $-\gamma_1 \gamma_2=\gamma_2\gamma_1$ correspond to the same bit string.
This bit string can equally be regarded as a vector in $\mathbb{F}_2^{\nphys}$. Given two operators $O_1,O_2$ with corresponding bit strings $b_1,b_2$, the product $O_1 O_2$ will correspond to the bit string $b_1 + b_2$ where addition is in $\mathbb{F}_2^{\nphys}$.
Thus, the stabilizer group is some subspace of $\mathbb{F}_2^{\nphys}$, with dimension $\nstab$.

So, Majorana fermion codes will correspond to subspaces of $\mathbb{F}_2^{\nphys}$ with the requirement
that the inner product of any two vectors in the subspace is equal to $0$; in the language of classical coding theory, these subspaces are self-orthogonal codes (note that in the case of Majoranas this subspace describes stabilizers while in the case of classical coding theory we interpret as a space of codewords).
To see this, note that the inner product of any vector in the subspace with itself must be zero (because stabilizers are products of an even number of Majorana operators).  The inner product of any two different vectors in the subspace must also be zero as follows: recall that any pair of operators $O_1,O_2$ in the stabilizer group must commute with each other.  We commute each operator $\gamma_a$ in $O_2$ through $O_1$ and keep track of the total sign; if $\gamma_a$ is also in $O_1$, then $\gamma_a$ anti-commutes with $O_1$ and otherwise it commutes (this follows because $O_1$ has an even number of Majorana operators and so if $\gamma_a$ is in $O_1$ then there are an odd number of operators in $O_1$ which anti-commute with $\gamma_a$); so, if $O_1,O_2$ commute then there are an even number of bits for the corresponding bit strings both contain a $1$.

The dimension of the code space is equal to
$$2^{\nphys/2-\nstab}.$$
We write
\be
\label{nlogis}
\nlog=\nphys/2-\nstab,
\ee
and we term $\nlog$ the number of ``logical qubits".

A logical operator is a product of Majorana operators  which commutes with all operators in the stabilizer group but which is not itself in the stabilizer group.
As shown in Ref.~\onlinecite{blt}, one can find $2\nlog$ logical operators $X_1,\ldots,X_\nlog,Z_1,\ldots,Z_\nlog$ which obey the usual Pauli commutation relations.  This motivates saying that there are $\nlog$ logical qubits.

\subsection{Distance}
The ``weight" of an operator is the Hamming weight of the corresponding bit string.
The distance of a code is defined to be the minimum of the weight of all nontrivial logical operators (here, nontrivial means not corresponding to the identity operator).

In this paper, we restrict to the case that the so-called ``fermion parity" operator $\gamma_1 \gamma_2 \ldots \gamma_{\nphys}$ is in the stabilizer group; when we refer to optimality properties of codes, we will always be considering this case, even though we will not state it from now on.
Thus, all logical operators must have even weight (otherwise, they would not commute with fermion parity) and so the distance of the code must be even and at least $2$.
There are two motivations for requiring that the fermion parity operator is in the stabilizer group.  First, physical implementations may naturally produce a code where it is in the stabilizer group due to charging energy effects\cite{Landau16,Plugge16a,Starkpatent16,Vijay16b,Plugge16b,karzig}.
Second, one cannot create superpositions of states without different fermion parity.  Conversely, in Ref.~\onlinecite{blt} it was suggested that codes with an odd weight logical operator might have better error correction properties by combining topological and parity protection.

Let $\nlm(\nphys,d)$ denote the maximal possible number of a logical qubits (over all possible codes) for a code with $\nphys$ physical Majorana modes and distance $d$.
A code with distance $d$ can detect any error acting on fewer than $d$ Majorana modes and can correct any error acting on fewer than $d/2$ Majorana modes.
Codes with distance $d=2$ are then not particularly useful: they cannot correct an error on a single Majorana mode; the simplest example of such a $d=2$ code is simply to take the stabilizer group to be generated by the fermion parity operator so that all even weight operators commute with the stabilizer group and no odd weight operators do.
In this paper, we will investigate some possible codes with small distance, $d=4$ and $d=6$.

The number $\nlm$ is non-decreasing in $\nphys$:
\be
\label{monotone}
\nlm(\nphys+2,d) \geq \nlm(\nphys,d).
\ee
To see this, consider a code $C$ with $\nphys$ physical Majoranas and $\nlog$ logical qubits.  Define a new code  $C'$for $\nphys+2$ physical Majoranas by taking the stabilizer group of $C'$ to be generated by the stabilizers of $C$ (acting on the first $\nphys$ operators out of the $\nphys+2$ physical Majoranas of $C'$) and also by the operator $\gamma_{\nphys+1}\gamma_{\nphys+2}$.  Then, any product of Majorana operators which commutes with the stabilizer group of $C'$ must be of the form $O$ or $O\gamma_{n_{phys}+1}\gamma_{n_{phys}+2}$ where $O$ is either a logical operator of $C$ or $O$ is in the stabilizer group of $C$.
If $O$ is not a logical operator for $C$ (i.e., $O$ is in the stabilizer group of $C$), then $O$ and $O\gamma_{n_{phys}+1}\gamma_{n_{phys}+2}$ are both in the stabilizer group of $C'$.
If $O$ is a logical operator, then it must have weight at least equal to the distance of $C$ and so $C'$ has the same distance as $C$.

Conversely, given a code $C'$ on $\nphys$ Majoranas which has an element of the stabilizer group with weight $2$, then such an 
element is equal to (after possibly relabelling the Majorana operators) $\gamma_{\nphys-1} \gamma_{\nphys}$ and $C'$ can be formed from a
code $C$ with $\nphys-2$ Majorana operators using the construction of the above paragraph.

A Majorana fermion code will be said to be ``degenerate" if there exists a nontrivial (i.e., not proportional to the identity) element of the stabilizer group with weight smaller than $d$, and it is said to be non-degenerate otherwise.

\subsection{Majorana Fermion Codes from Qubit Stabilizer Codes}
\label{mapping}
We very briefly review the construction of Majorana fermion codes given a qubit stabilizer code.  Given
a qubit stabilizer code with $\nqub$ qubits, construct a Majorana fermion code with $\nphys=4\nqub$ as follows.
For each qubit $i$ of the qubit stabilizer code, define $4$ Majorana fermions labelled by a pair $(i,a)$, where $a\in \{1,\ldots,4\}$.
The stabilizer group of the Majorana fermion code is generated by the following stabilizers.  First, for every $i$,
we have the stabilizer $\gamma_{(i,1)}\gamma_{(i,2)}\gamma_{(i,3)}\gamma_{(i,4)}$.  The four Majorana fermions $i,a$ have a four dimensional Hilbert space, but the $+1$ eigenspace of this stabilizer is only two dimensional and so corresponds to a qubit.  Then, we can identify the operator $X_i$ in the qubit code with $\gamma_{(i,1)}\gamma_{(i,2)}$ and identify the operator $Z_i$ with $\gamma_{(i,1)} \gamma_{(i,3)}$.
Then, for every stabilizer of the qubit code, we map that stabilizer to a stabilizer of the Majorana fermion code, by replacing each $X_i$ or $Z_i$ with the appropriate $\gamma_{(i,1)}\gamma_{(i,2)}$ or $\gamma_{(i,1)}\gamma_{(i,3)}$, respectively.

As shown in Refs.~\onlinecite{blt,anyons}, the distance of the resulting Majorana fermion code is twice the distance of the qubit stabilizer code.

\section{Distance $d=4$ Codes: Analytical Results}
In this section, we consider codes with distance $d=4$, and give some analytical results.  In the next section
we give the results of a numerical search for $d=4$ codes as well as $d=6$ codes.

\subsection{Upper Bounds}
Let $\nlm^{d}(\nphys,d)$ denote the maximal number of logical qubits for a degenerate code with distance $\nphys$ physical qubits and distance $d$ and
let $\nlm^{nd}(\nphys,d)$ denote the maximal number of logical qubits for a non-degenerate code with $\nphys$ physical qubits and distance $d$.
For a degenerate code with $d=4$, there must be an element of the stabilizer group with weight $2$, and so, by the discussion below Eq.~\ref{monotone}, we have $\nlm^d(\nphys,4)=\nlm(\nphys-2,4)$.  If, in turn, $\nlm(\nphys-2,4)=\nlm^d(\nphys-2,4)$ then $\nlm(\nphys-2,4)=\nlm(\nphys-4,4)$.
Proceeding in this fashion, we find that
\be
\label{red}
\nlm^{d}(\nphys,4)={\rm max}_{\stackrel{0\leq M<\nphys}{M \; {\rm even}}} \;\nlm^{nd}(M,4).
\ee
Further,
\be
\label{max}
\nlm(\nphys,4)={\rm max}\Bigl(\nlm^{nd}(\nphys,4),\nlm^d(\nphys,4)\Bigr).
\ee
and so 
\be
\label{max2}
\nlm(\nphys,4)={\rm max}_{\stackrel{0\leq M\leq \nphys}{M \; {\rm even}}} \; \nlm^{nd}(M,4).
\ee
Thus, it suffices to determine $\nlm^{nd}(M,4)$ for all $M \leq \nphys$ in order to determine $\nlm(\nphys,4)$ and so that is what we now consider.

Further, we claim that
\be
\label{ceil}
\nlm^{nd}(\nphys,4)\leq \nphys/2-\lceil \log_2(\nphys) \rceil -1.
\ee
To see this, note that in a non-degenerate code, any nontrivial operator with weight $t<d$ will fail to commute with at least one stabilizer.
The set of stabilizer generators that the operator does not commute with is called the ``error syndrome".
This set can be written as a bit string of length $\nstab$, or, equivalently, as a vector in $\mathbb{F}_2^{\nstab}$.
For $d=4$, this means that any operator $\gamma_a \gamma_b$ with $a\neq b$ has weight $t=2<d=4$ and so this operator has a nontrivial error syndrome (here, nontrivial means the syndrome includes at least one generator). 
Further, the operators $\gamma_a$ and $\gamma_b$ also have nontrivial error syndromes and so
because $\gamma_a \gamma_b$ has a nontrivial error syndrome, the error syndromes of $\gamma_a$ and $\gamma_b$ must be distinct because
the error syndrome of the product of two operators is simply the sum of the error syndromes, viewed as vectors in
$\mathbb{F}_2^{\nphys}$. Thus, 
each single Majorana operator $\gamma_a$ for $a\in \{1,\ldots,\nphys\}$ must correspond to a unique error syndrome.  
There are $\nstab$ generators and hence $2^{\nstab-1}$ nontrivial error syndromes.  However,
one generator is fermion parity and all single Majorana operators $\gamma_a$ anti-commute with this operator and hence there are only $2^{\nstab-1}$ possible error syndromes for a single Majorana operator (the error syndrome will always be nontrivial since any single Majorana operator anti-commutes with fermion parity).
Hence, for a non-degenerate code,
$2^{\nstab-1} \geq \nphys$; using Eq.~(\ref{nlogis}), this implies Eq.~(\ref{ceil}).

Another way to see that each single Majorana operator must correspond to a unique error syndrome is that
single Majorana operator errors are correctable and so for a non-degenerate code it must be possible to determine the error from the syndrome.

Given Eq.~(\ref{ceil}) and Eq.~(\ref{max2}), it follows that
\be
\label{ceil2}
\nlm(\nphys,4)\leq \nphys/2-\lceil \log_2(\nphys) \rceil -1.
\ee

\subsection{``Hamming Majorana Codes" with $\nphys=2^\li$}
Naively, one might think that for any $\nphys$ one can construct a code with a $\nlog$ that saturates this inequality (\ref{ceil2}).  After all, it would seem that one could always choose the stabilizers
such that the error syndrome gives enough information to uniquely identify any single Majoran fermion error, i.e., to ensure that each single Majorana fermion operator has a unique error syndrome.  However, the constraints that the stabilizers must be bosonic and must commute with each other may make this impossible for some $\nphys$.
In this subsection, we show that the inequality is saturated for the particular case that $\nphys$ is a power of $2$:
\be
\li \geq 3 \; \rightarrow \; \nlm(2^\li,4)=\nlm^{nd}(2^\li,4)=2^{\li-1}-\li-1.
\ee
For $\li=3$, the construction of this section will give a code with $\nlog=0$ but which has a unique syndrome for each single Majorana error.

The class of codes we construct is closely related to the classical Hamming code, so we call them ``Hamming Majorana codes".

We generate the stabilizer group by $\li$ different stabilizers, labelled $S_1,S_2,\ldots,S_\li$, and by the fermion parity operator, so that $\nstab=\li+1$.
The stabilizer $S_m$ will be the product of all operators $\gamma_a$ such that the $m$-th bit of $a-1$ in binary is equal to $1$ (we count the $m$-th bit from the right, so that the first bit is the least significant, and so on; the order in which we count it is completely arbitrary but we choose to count from the right as it makes the matrix below look nicer).
Note that $a-1$ ranges from $0$ to $\nphys-1$.
Thus, in the case $m=4$, consider the following matrix:
$$S=
\begin{pmatrix}
0000\\
0001\\
0010\\
0011\\
0100\\
0101\\
0110\\
0111\\
1000\\
1001\\
1010\\
1011\\
1100\\
1101\\
1110\\
1111\\
\end{pmatrix}.
$$
The matrix
$S$ is a $16$-by-$4$ matrix.  The rows of $X$ label different Majorana operators and the columns of $m$ label different stabilizers, $S_1,...,S_4$.
So, for example, the operator $\gamma_{13}$ corresponds to the $12$-th row of this table; in binary, $12$ is $1100$ and so stabilizers $S_3$ and $S_4$ include operator $\gamma_{13}$.

It is clear that such a choice of stabilizers gives each $\gamma_a$ a unique error syndrome.  
Indeed, the pattern of violated stabilizers is given by the binary representation of $a$. 
Further, each $S_i$ has weight $2^{\li-1}$ and so has even weight for $\li\geq 2$.  Also, given any pair $S_i,S_j$ for $i\neq j$, the number of operators $\gamma_a$ which are in both $S_i$ and $S_j$ is equal to $2^{\li-2}$ and so is even for $\li\geq 3$.
So, for $\li\geq 3$, this defines a valid code.

\begin{figure}
\includegraphics[width=3in]{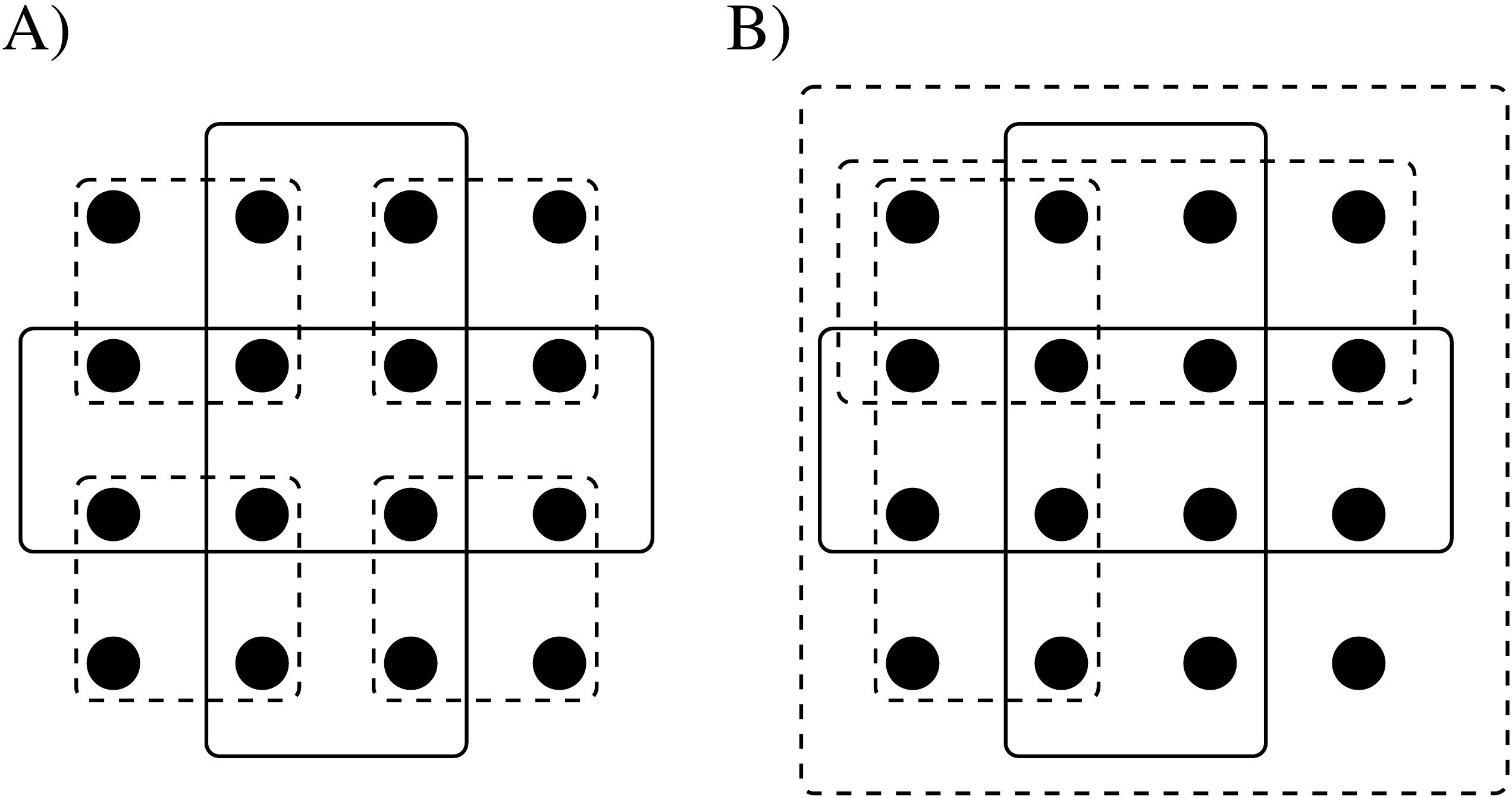}
\caption{(A) shows stabilizers for a $16$ Majorana code derived from the $4$ qubit code.  Each circle represents a Majorana mode.  Each rounded rectangle (surrounded by either a solid or a dashed line) represents a stabilizer; the stabilizer is the product of Majorana operators on the modes contained inside that rectangle.  There are $6$ independent stabilizers and hence $\nlog=2$. 
The rounded rectangles with dashed lines indicate generators acting on $4$ Majorana operators; these are the stabilizers $\gamma_{i,1} \gamma_{i,2} \gamma_{i,3} \gamma_{i,4}$ of the mapping of subsection \ref{mapping}.
(B) shows stabilizers for the Hamming Majorana code with $\nphys=16$.  There are now only $\nstab=5$ independent stabilizers.  One of the dashed rounded rectangles in (B) surrounds all qubits; this is the parity operator.
The solid rounded rectangles are the same in (B) as in (A).  The dashed rounded rectangles in (B) generate a subgroup of the dashed rounded rectangles in (A).}
\label{16fig}
  \end{figure}

For $\li=4$, this defines a code $\nphys=16,\nlog=3,d=4$.  It is interesting to compare this to another code on $16$ Majorana fermions.  There is a $4$ qubit code with distance $2$ and $2$ logical qubits with
stabilizers $X_1 X_2 X_3 X_4$ and $Z_1 Z_2 Z_3 Z_4$.
Applying the mapping of subsection \ref{mapping} to this $4$ qubit code
gives a Majorana fermion code with $\nphys=16,\nlog=2,d=4$.  
Fig.~\ref{16fig} shows the stabilizers for these two codes.

Note also that there is no qubit stabilizer code on $4$ physical qubits with distance $2$ and $3$ logical qubits (we leave this as an exercise for the reader or see Ref.~\onlinecite{codetables}).
Hence, no Majorana fermion code with $\nphys=16$ derived by mapping from a qubit stabilizer code has as many logical qubits as the Hamming Majorana code.

 Eq.~\ref{ceil} shows that no code with $\nphys\leq 10$ and distance $d=4$ can have $\nlog>0$.  We show in the next paragraph that no code with $\nphys=12$ has $\nlog>0$.
 For $\nphys=14$,  we performed a numerical search (described in the next section) for a code with $\nstab=6$ and did not succeed.  Hence, we believe that no such code exists, i.e., we believe that $\nphys=16$ is the minimum number of modes to have $d=4,\nlog>0$.
 
 For $\nphys=12$, in order to have $\nlog>0$ we must have $\nstab=5$.  
 We now show that this is not possible.
 One of these generators is the fermion parity operator.  Call the other generators $g_1,g_2,g_3,g_4$.  Of the other stabilizers, there must be one (which we will call $g_4$) with weight $4$ (proof: stabilizers have event weight, so possible nontrivial weights are $2,4,6,8,10$.  We can multiply a stabilizer of weight $w$ by the fermion parity operator to give a stabilizer with weight $12-w$.  So, we can assume the generators have weights $2,4,6$.  For a non-degenerate code, no generators have weight $2$, so we can take generators to have weights $4,6$.  Given two distinct generators with weights $6$ (such that their product is not the fermion parity operator), their product must be $0$ mod $4$ (since they commute, so there are an even number of modes that they both act on), so we can assume the product has weight $4$).  Given a stabilizer of weight $4$, without loss of generality let it be $\gamma_9 \gamma_{10} \gamma_{11} \gamma_{12}$.  There are $8$ possible single Majorana errors which commute with this stabilizer (errors on Majorana modes $\gamma_1,\ldots,\gamma_{8}$), so  we need the remaining $3$ generators to uniquely identify those errors.  The only way to uniquely identify those errors (since there are $8$ possible errors and $2^3$ syndromes) is to use something similar to a Hamming Majorana code: the remaining $3$ stabilizer generators $g_1,g_2,g_3$ must each be a generator of the Hamming Majorana code with $\nphys=8$ on the first $8$ modes multiplied by some product of $\gamma_9,\ldots,\gamma_{12}$.  Call these products $p_1,p_2,p_3$, respectively; i.e., $g_a$ is equal to a Hamming Majorana generator on the first $8$ modes multiplied by $p_a$.  Since $g_1,g_2,g_3$ commute with each other, the operators $p_1,p_2,p_3$ commute with each other, and further all have even weight.  Hence, up to multiplication by $g_4$, and up to permutation of modes $9,10,11,12$, the operators $p_a$ are equal to either identity or $\gamma_9 \gamma_{10}$.  Hence, we cannot have a unique syndrome for each single Majorana error.

\section{Numerical Search For Other Codes}
We now describe a numerical search for other codes with $d=4,6$.  We begin with the case $d=4$, and describe the algorithm there.  This algorithm is based on a random walk through codes.  We then describe some properties of the walk.  Finally, we discuss modifications to the algorithm for the case $d\geq 6$ and give results for $d=6$.

\subsection{Distance $d=4$ Codes}
The Hamming Majorana codes give distance $4$ codes with optimal $\nlog$ for $\nphys=16,32$.
For other values of $\nstab$ with $18\leq \nstab \leq 30$, we conducted a numerical search for other distance $d=4$ codes.
We searched only for non-degenerate codes.

The search was a random search, implemented as follows.
We choose given values of $\nphys$ and $\nstab$.  The algorithm searches through codes until either it finds that a distance $4$ code or until it gives up after a sufficiently large number of iterations.  One stabilizer generator will
be the fermion parity operator, which is not explicitly stored, so in fact the algorithm only stores the remaining $\nstab-1$ generators as the way it defines the code.  We refer to these $\nstab-1$ generators as the ``stored list".

We initialize the stored list to
to $\gamma_1 \gamma_2$ and $\gamma_3 \gamma_4$ and so on, up to $\gamma_{2(\nstab-1)-1}\gamma_{2(\nstab-1)}$.  In addition, there is the fermion parity operator, as mentioned above.
This is a valid code (in that all stabilizers have even weight and commute with each other) but it has only distance $2$.

Then, the algorithm iterates the following for some number of steps.  First, it randomly updates the stabilizers.
This is done by choosing $4$ different Majorana modes at random.  Let these modes be $i,j,k,l$.  Then,
it performs the replacements:
\begin{eqnarray}
\label{replacement}
\gamma_i &\rightarrow &\gamma_j \gamma_k \gamma_l, \\
\gamma_j &\rightarrow &\gamma_i \gamma_k \gamma_l, \\
\gamma_k &\rightarrow &\gamma_i \gamma_j \gamma_l, \\
\gamma_l &\rightarrow &\gamma_i \gamma_j \gamma_k.
\end{eqnarray}
That is, for each stabilizer generator in the stored list, it replaces every occurrence of $\gamma_i$ by $\gamma_j \gamma_k \gamma_l$.  These replacements are all performed in parallel; that is, $\gamma_i \gamma_j$ is replaced by $\gamma_j \gamma_k \gamma_l \gamma_i \gamma_k \gamma_l$.
Note that we do not care about the sign of the stabilizer generator (different choices of signs define a code with the same $d,\nstab$), so we do not track the sign during this replacement.
Note also that this replacement does not change the fermion parity operator.

This update procedure allows us to perform a rapid random walk through different codes.  The advantage of doing it this way is that each time we generate a new code, we are guaranteed that it will be valid, having even weight stabilizers that commute with each other, as the replacements maintain the algebra of anti-commutation relations obeyed by the Majorana operators.

An alternative way to define the replacement is that if a stabilizer generator contains an {\it odd} number of operators $\gamma_i,\gamma_j,\gamma_k,\gamma_l$ then that generator is multiplied by $\gamma_i \gamma_j \gamma_k \gamma_l$, up to signs.
This update can be performed very quickly using bitwise operations, storing each stabilizer as a bit string, then ANDing the bit string with a mask which is a $1$ in the bits corresponding to $i,j,k,l$ (we pre-compute these masks for all ${N \choose 4}$ choices of $i<j<k<l$) and then count the number of $1$ bits; if this number is odd, we XOR the bit string with the mask.

Then, once the new code is generated, we check if it has distance $4$.  This can again be done with bitwise operations.  For of the ${N \choose 2}$ different operators $\gamma_i \gamma_j$ with $i<j$, we generate a mask with a $1$ in the bits corresponding to $i,j$.  We then check whether, for each mask, there is at least one stabilizer generator which anti-commutes; if so, the code has distance $d>2$.  This can be done by ANDing the mask with the bit string corresponding to the generator and counting if there are an odd number of $1$s in the result.  If we find a code with distance $d>2$ we report success, otherwise we continue.

For each $\nphys$, we tried increasing values of $\nstab$ until we found a code.  For each value of $\nstab$ we did $2000$ independent runs with $10^8$ steps on each run.  Only if all those runs failed did we increase the value of $\nstab$ and try again.
The results are shown in Table \ref{nondegentable}.  This gives the best non-degenerate codes found; note that $\nlog$ is non-monotonic with $\nphys$.
Using this table and Eq.~(\ref{red}), we give the best codes in table \ref{degend2table}, and also compare to the best codes derived from qubit stabilizer codes using the mapping of subsection \ref{mapping}.
Generators for some of the codes are given in the Appendix.

As a test of the algorithm, we also ran it for $\nphys=32,\nstab=6$, where on $882$ out of the $2000$ runs it succeeded in finding a code with distance $4$.  We know such a code exists (the Hamming Majorana code), so this gives some indication that the algorithm will find a code when it exists.

\begin{table}
\begin{tabular}{c|c|c|c}
$\nphys$ & $\nstab$ & $\nlog$  & \\
16 & {5} & {3} & \checkmark\\
18 & 7 & 2\\
{20} & {6} & {4} & \checkmark\\
22 & 7 & 4 \\
{24} & {6} & {6} & \checkmark\\
26 & 7 & 6 \\
{28} & {7} & {7} & \checkmark \\
{30} & {7} & {8} &\checkmark
\end{tabular}
\caption{Table showing non-degenerate codes found for $\nphys=16,\ldots,30$.  
The value for $16$ is the Hamming Majorana code, others are from computer search as explained in text.
The value $\nstab$ is the
smallest value of $\nstab$ for the given $\nphys$ for which we found a non-degenerate distance $2$ code.
Lines with a checkmark indicate that that code has larger $\nlog$ than any code we found with smaller $\nphys$; these lines are used to make table \ref{degend2table}.}
\label{nondegentable}
\end{table}

\begin{table}
\begin{tabular}{c|c|c|c}
$\nphys$ & $\nstab$ & $\nlog$  & $\nlog_{qubit}$\\
16 & {5} & {3} & 2\\
18 & 6 & 3\\
{20} & {6} & {4}& 2\\
22 & 7 & 4 \\
{24} & {6} & {6} &4 \\
26 & 7 & 6 \\
{28} & {7} & {7} &4 \\
{30} & {7} & {8} \\
{32} & 6 & 10 &6
\end{tabular}
\caption{Table showing codes found for $\nphys=16,\ldots,32$, including both degenerate and non-degenerate codes.
The table is built using the codes with a checkmark in Table \ref{nondegentable} and using Eq.~(\ref{red}).
The final column, called $\nlog_{qubit}$ and given only for codes where $\nphys$ is a multiple of $4$, is the maximum possible number of logical qubits for a distance $4$ Majorana fermion code derived from a distance $2$ qubit stabilizer code; we use the bounds from Ref.~\onlinecite{codetables} to get $\nlog_{qubit}$ for the qubit stabilizer codes.
}
\label{degend2table}
\end{table}

\subsection{Properties of Random Walk}
\label{prop}
There are two important properties of the random walk described above.  First, the transition probabilities obey detailed balance as follows.  Let $c$ represent the state of the algorithm, namely the stored list of stabilizers.  Let $P_{c,c'}$ denote the transition probability from state $c$ to some other state $c'$.  Then, these probabilities obey detailed balance in that $P_{c,c'}=P_{c',c}$.  To see this, note that if some given choice of $i,j,k,l$ leads to a transition from $c$ to $c'$, then the same choice of $i,j,k,l$ leads to a transition from $c'$ to $c$.

Second, consider any stored list $c$ such that the list has $\nstab-1$ independent stabilizers, and such that the fermion parity operator is not in the group generated by the stored list of stabilizers.  We will show that there is a sequence of replacements that turns this stored list into the list of stabilizers 
$\gamma_1 \gamma_2, \gamma_3 \gamma_4, \ldots, \gamma_{2 (\nstab-1)-1} \gamma_{2(\nstab-1)}$, up to possibly a permutation of the Majorana operators.  Combined with detailed balance above, this implies that, up to a permutation of the Majorana operators, the random walk will ultimately explore all possible codes with the given $\nstab$ up to permutation of the Majorana operators.

Consider the first stabilizer in the list.  We show how to turn it into $\gamma_1 \gamma_2$.  The stabilizer cannot have weight $\nphys$ since it is not equal to the fermion parity operator.  If it has weight between $4$ and $\nphys-2$, we can find $i<j<k$ such that the stabilizer includes $\gamma_i,\gamma_j, \gamma_k$ and we can find an $l$ such that the stabilizer does not include $\gamma_l$.  Performing the replacement with the given $i,j,k,l$
reduces the weight of the stabilizer by $2$.  Continue in this fashion until it has weight $2$.
Once has weight $2$, then we can turn it into $\gamma_1 \gamma_2$ by permutations.  

We now repeat the procedure for the second stabilizer in the list, but we only consider the action of the stabilizer on modes $\gamma_3,\ldots,\gamma_{\nphys}$.  That is, we ignore the bits in the bit string corresponding to modes $1,2$, and define the ``weight" to be the number of other bits which are nonzero.  The weight of the stabilizer must be less than $\nphys-2$ since the fermion parity operator is not in the group generated by the first two stabilizers.
We find $i,j,k,l$ as in the above paragraph, choosing $2<i,j,k,l$, reducing the weight of the stabilizer until it is equal to $2$.  Then, once the weight is equal to $2$, we permute until the stabilizer is equal to $\gamma_3 \gamma_4$, possibly multiplied by $\gamma_1 \gamma_2$.

We continue this procedure for the third, fourth, etc... stabilizers.  On the $j$-th stabilizer, we ignore the first $2(j-1)$ bits in the bit string, and reduce the weight of the remaining bits.  We then apply permutations until the stabilizer is equal to $\gamma_{2j-1} \gamma_{2j}$, possibly multiplied by earlier stabilizers in the list.

This procedure required using permutations.  If $\nphys\geq 5$ (as it is in all cases of interest), the group generated by Eq.~(\ref{replacement}) includes permutations, so in fact the random walk explores all possible permutations.  To see that the group includes permutations in this case, consider five modes, $\gamma_1,\ldots,\gamma_5$.  Apply Eq.~(\ref{replacement}) three times, using first $i=1,j=2,k=3,l=4$, then $i=2,j=3,k=4,l=5$ and finally $i=1,j=2,k=3,l=4$.  Then, up to signs, the effect is to map $\gamma_1 \leftrightarrow \gamma_5$, while preserving $\gamma_2,\gamma_3,\gamma_4$.  Since any exchange of a pair of Majoranas is in the group, the group contains all permutations.

\subsection{Distance $d=6$ Codes}
\label{d6sec}
We also performed a numerical search for codes with distance $d=6$.  In this case, we searched for all possible codes, degenerate or not.  We used a similar algorithm to the search for $d=4$ codes.
We initialized the stored list of generators in the same way as in the search for $d=4$.  We used the same Eq.~(\ref{replacement}) to update stabilizer generators to perform a random walk through codes.
However, we also store a set of generators for $2\nlog$ independent logical operators.  These are initialized to logical operators of the initial code, and then are also updated using Eq.~(\ref{replacement}).  This list is used in checking distance of the code.

The only change is how we tested the distance of the code.  Since we are looking for a code with distance $6$, we need to check operators with weight $4$ as well as those with weight $2$, and since we might be including degenerate codes, we need to check if there is an operator of weight $2$ or $4$ which commutes with all generators {\it and} which does not commute with at least one logical operator.  Checking that an operator of weight $2$ or $4$ commutes with a generator is done in the same way as in the search for $d=4$ codes (we store a mask for each such operator, and we AND the mask we each generator and count the bits in the result).  To check commutation with logical operators, we use the list of logical operators of the code that we have stored and again use bitwise operations. 

We used the algorithm in the same way, choosing a given $\nphys$ and increasing $\nstab$ until a $d=6$ code was found.  As before, we ran the algorithm $2000$ times, taking $10^8$ steps for each run, until giving up and increasing $\nstab$.  The results are shown in Table \ref{6table}, as well as a comparison to the best distance $6$ Majorana fermion codes derived from a qubit stabilizer code.
Generators for these codes are shown in the Appendix.

The code with $\nphys=20$ has the same number of qubits as a code derived from a qubit stabilizer code.  Indeed, the Majorana fermion code that we found (at least for all runs we inspected) was a code derived from a qubit stabilizer code.  For $\nphys=28$, the code has more logical qubits than a code derived from a qubit stabilizer code.  We found in this case (at least for all runs that we inspected) that the code had one weight $4$ stabilizer in the stabilizer group; thus, one may also build such a code out of $1$ qubit and $24$ Majorana fermions.  With $\nphys=30$, for all runs that we inspected, there were no weight $4$ stabilizers in the stabilizer group.

\begin{table}
\begin{tabular}{c|c|c|c|c}
$\nphys$ & $\nstab$ & $\nlog$  & $\nlog_{qubit}$ \\
20 & 9 & 1 &1\\
28 & 12 & 2 &1\\
30 & 12 & 3
\end{tabular}
\caption{Table showing optimal codes found for $\nphys\leq 32$.  If an entry is not present in the table for a given $\nphys$, it means that the best code found for that $\nphys$ had the same $\nlog$ as a code in the table with a smaller $\nphys$.  For example, the optimal code found with $\nphys=32$ has $\nstab=13$ and hence $\nlog=3$, the same as the code shown in the table with $\nphys=30$.  No distance $d=6$ codes were found with $\nphys<20$.
The column $\nlog_{qubit}$, given only for codes where $\nphys$ is a multiple of $4$, gives the maximum possible number of logical qubits for a distance $6$ Majorana fermion code derived from a distance $3$ qubit stabilizer code; we use the bounds from Ref.~\onlinecite{codetables} to get $\nlog_{qubit}$ for the qubit stabilizer codes.}
\label{6table}
\end{table}

\section{Discussion and Implementation}
We have given several small Majorana fermion codes.  Interestingly, there exist codes whose performance is better than that of any code derived from a qubit code.  The simplest one, the Hamming Majorana code with $\nphys=16$ in fact has a stabilizer group which is a subgroup of the stabilizer group of the Majorana code derived from a $4$ qubit code.

We have also given a numerical search algorithm.  Using bitwise operations, this search can be run extremely quickly.  It is not exhaustive, so the failure of the algorithm does not prove the non-existence of a code, but we believe that the codes we have found are optimal.
It may be possible to run a similar numerical search for qubit codes.  The basic idea of the random search is that it allows us to turn a valid set of stabilizers (obeying commutation relations) to another valid set; one could construct a similar search algorithm for qubit stabilizer codes by randomly applying operations from the Clifford group.

Efficient implementation of this code can be most easily be done if it is possible to measure the stabilizers directly, as in the scheme of Ref.~\onlinecite{hyart}.  One property of these codes is that each stabilizer (except for the fermion parity operator) can be written in two different, non-overlapping ways.  For example, with $\nphys=16$, the operators $\gamma_1 \ldots \gamma_8$ and $\gamma_9 \ldots \gamma_{16}$ agree, up to fermion parity.  Hence, this provides two distinct ways to measure the same stabilizer; these independent measurements may allow one to reduce the effect of measurement errors.  The $\nphys=16$ Hamming Majorana code has a physical layout, shown in Fig.~\ref{16fig}, which may simplify some of these measurements, as the generators are all contained in local regions (squares or rectangles).

{\it Acknowledgments---} I thank D. Wecker for useful discussions.
\bibliography{mcode}

\begin{thebibliography}{11}
\expandafter\ifx\csname natexlab\endcsname\relax\def\natexlab#1{#1}\fi
\expandafter\ifx\csname bibnamefont\endcsname\relax
  \def\bibnamefont#1{#1}\fi
\expandafter\ifx\csname bibfnamefont\endcsname\relax
  \def\bibfnamefont#1{#1}\fi
\expandafter\ifx\csname citenamefont\endcsname\relax
  \def\citenamefont#1{#1}\fi
\expandafter\ifx\csname url\endcsname\relax
  \def\url#1{\texttt{#1}}\fi
\expandafter\ifx\csname urlprefix\endcsname\relax\def\urlprefix{URL }\fi
\providecommand{\bibinfo}[2]{#2}
\providecommand{\eprint}[2][]{\url{#2}}

\bibitem[{\citenamefont{Gottesman}(1997)}]{stab}
\bibinfo{author}{\bibfnamefont{D.}~\bibnamefont{Gottesman}},
  \bibinfo{journal}{arXiv preprint quant-ph/9705052}  (\bibinfo{year}{1997}).

\bibitem[{\citenamefont{Bravyi et~al.}(2010)\citenamefont{Bravyi, Terhal, and
  Leemhuis}}]{blt}
\bibinfo{author}{\bibfnamefont{S.}~\bibnamefont{Bravyi}},
  \bibinfo{author}{\bibfnamefont{B.~M.} \bibnamefont{Terhal}},
  \bibnamefont{and} \bibinfo{author}{\bibfnamefont{B.}~\bibnamefont{Leemhuis}},
  \bibinfo{journal}{New Journal of Physics} \textbf{\bibinfo{volume}{12}},
  \bibinfo{pages}{083039} (\bibinfo{year}{2010}).

\bibitem[{\citenamefont{Kitaev}(2006)}]{anyons}
\bibinfo{author}{\bibfnamefont{A.}~\bibnamefont{Kitaev}},
  \bibinfo{journal}{Annals of Physics} \textbf{\bibinfo{volume}{321}},
  \bibinfo{pages}{2} (\bibinfo{year}{2006}).

\bibitem[{cod()}]{codetables}
\urlprefix\url{http://www.codetables.de}.

\bibitem[{\citenamefont{Karzig et~al.}(2016)\citenamefont{Karzig, Knapp,
  Lutchyn, Bonderson, Hastings, Nayak, Alicea, Flensberg, Plugge, Oreg
  et~al.}}]{karzig}
\bibinfo{author}{\bibfnamefont{T.}~\bibnamefont{Karzig}},
  \bibinfo{author}{\bibfnamefont{C.}~\bibnamefont{Knapp}},
  \bibinfo{author}{\bibfnamefont{R.}~\bibnamefont{Lutchyn}},
  \bibinfo{author}{\bibfnamefont{P.}~\bibnamefont{Bonderson}},
  \bibinfo{author}{\bibfnamefont{M.}~\bibnamefont{Hastings}},
  \bibinfo{author}{\bibfnamefont{C.}~\bibnamefont{Nayak}},
  \bibinfo{author}{\bibfnamefont{J.}~\bibnamefont{Alicea}},
  \bibinfo{author}{\bibfnamefont{K.}~\bibnamefont{Flensberg}},
  \bibinfo{author}{\bibfnamefont{S.}~\bibnamefont{Plugge}},
  \bibinfo{author}{\bibfnamefont{Y.}~\bibnamefont{Oreg}}, \bibnamefont{et~al.},
  \bibinfo{journal}{arXiv preprint arXiv:1610.05289}  (\bibinfo{year}{2016}).

\bibitem[{\citenamefont{{Landau} et~al.}(2016)\citenamefont{{Landau}, {Plugge},
  {Sela}, {Altland}, {Albrecht}, and {Egger}}}]{Landau16}
\bibinfo{author}{\bibfnamefont{L.~A.} \bibnamefont{{Landau}}},
  \bibinfo{author}{\bibfnamefont{S.}~\bibnamefont{{Plugge}}},
  \bibinfo{author}{\bibfnamefont{E.}~\bibnamefont{{Sela}}},
  \bibinfo{author}{\bibfnamefont{A.}~\bibnamefont{{Altland}}},
  \bibinfo{author}{\bibfnamefont{S.~M.} \bibnamefont{{Albrecht}}},
  \bibnamefont{and} \bibinfo{author}{\bibfnamefont{R.}~\bibnamefont{{Egger}}},
  \bibinfo{journal}{Physical Review Letters} \textbf{\bibinfo{volume}{116}},
  \bibinfo{eid}{050501} (\bibinfo{year}{2016}), \eprint{1509.05345}.

\bibitem[{\citenamefont{{Plugge} et~al.}({\natexlab{a}})\citenamefont{{Plugge},
  {Landau}, {Sela}, {Altland}, {Flensberg}, and {Egger}}}]{Plugge16a}
\bibinfo{author}{\bibfnamefont{S.}~\bibnamefont{{Plugge}}},
  \bibinfo{author}{\bibfnamefont{L.~A.} \bibnamefont{{Landau}}},
  \bibinfo{author}{\bibfnamefont{E.}~\bibnamefont{{Sela}}},
  \bibinfo{author}{\bibfnamefont{A.}~\bibnamefont{{Altland}}},
  \bibinfo{author}{\bibfnamefont{K.}~\bibnamefont{{Flensberg}}},
  \bibnamefont{and} \bibinfo{author}{\bibfnamefont{R.}~\bibnamefont{{Egger}}},
  \emph{\bibinfo{title}{{Roadmap to Majorana surface codes}}},
  \bibinfo{note}{arXiv:1606.08408}.

\bibitem[{\citenamefont{Lutchyn et~al.}()\citenamefont{Lutchyn, Alicea,
  Bonderson, Freedman, Karzig, Knapp, and Nayak}}]{Starkpatent16}
\bibinfo{author}{\bibfnamefont{R.}~\bibnamefont{Lutchyn}},
  \bibinfo{author}{\bibfnamefont{J.}~\bibnamefont{Alicea}},
  \bibinfo{author}{\bibfnamefont{P.}~\bibnamefont{Bonderson}},
  \bibinfo{author}{\bibfnamefont{M.}~\bibnamefont{Freedman}},
  \bibinfo{author}{\bibfnamefont{T.}~\bibnamefont{Karzig}},
  \bibinfo{author}{\bibfnamefont{C.}~\bibnamefont{Knapp}}, \bibnamefont{and}
  \bibinfo{author}{\bibfnamefont{C.}~\bibnamefont{Nayak}},
  \emph{\bibinfo{title}{{Measuring and manipulating Majorana quasiparticle
  states using the Stark effect}}}, \bibinfo{note}{08-22-2016, U.S. Provisional
  application, pending}.

\bibitem[{\citenamefont{{Vijay} and {Fu}}(2016)}]{Vijay16b}
\bibinfo{author}{\bibfnamefont{S.}~\bibnamefont{{Vijay}}} \bibnamefont{and}
  \bibinfo{author}{\bibfnamefont{L.}~\bibnamefont{{Fu}}},
  \emph{\bibinfo{title}{{Braiding without Braiding: Teleportation-Based Quantum
  Information Processing with Majorana Zero Modes}}} (\bibinfo{year}{2016}),
  \bibinfo{note}{arXiv:1609.00950}.

\bibitem[{\citenamefont{{Plugge} et~al.}({\natexlab{b}})\citenamefont{{Plugge},
  {Rasmussen}, {Egger}, and {Flensberg}}}]{Plugge16b}
\bibinfo{author}{\bibfnamefont{S.}~\bibnamefont{{Plugge}}},
  \bibinfo{author}{\bibfnamefont{A.}~\bibnamefont{{Rasmussen}}},
  \bibinfo{author}{\bibfnamefont{R.}~\bibnamefont{{Egger}}}, \bibnamefont{and}
  \bibinfo{author}{\bibfnamefont{K.}~\bibnamefont{{Flensberg}}},
  \emph{\bibinfo{title}{{Majorana box qubits}}},
  \bibinfo{note}{arXiv:1609.01697}.

\bibitem[{\citenamefont{Hyart et~al.}(2013)\citenamefont{Hyart, van Heck,
  Fulga, Burrello, Akhmerov, and Beenakker}}]{hyart}
\bibinfo{author}{\bibfnamefont{T.}~\bibnamefont{Hyart}},
  \bibinfo{author}{\bibfnamefont{B.}~\bibnamefont{van Heck}},
  \bibinfo{author}{\bibfnamefont{I.}~\bibnamefont{Fulga}},
  \bibinfo{author}{\bibfnamefont{M.}~\bibnamefont{Burrello}},
  \bibinfo{author}{\bibfnamefont{A.}~\bibnamefont{Akhmerov}}, \bibnamefont{and}
  \bibinfo{author}{\bibfnamefont{C.}~\bibnamefont{Beenakker}},
  \bibinfo{journal}{Physical Review B} \textbf{\bibinfo{volume}{88}},
  \bibinfo{pages}{035121} (\bibinfo{year}{2013}).

\end{thebibliography}
\newpage

\appendix
\section{Table of Codes Found For $d=4$}
\label{d4section}

We give in Table \ref{d4gentable} a table of some of the codes found using numerical search for distance $d=4$.
We show stabilizers for non-degenerate codes with $\nphys=20,24,28,30$.  These are the codes shown with a checkmark in table \ref{nondegentable}.

\begin{table}[hbtp]
\begin{tabular}{l}
$\nphys=20$\\
01001101010001011101\\
10011010110101111100\\
11010110101011000100\\
01101010101100101001\\
10100111001010111101\\
\hline
$\nphys=24$\\
110110110100100010101001\\
000010110111010010010111\\
111000100000001011110111\\
101000011001000000001011\\
001111100001011000101101\\
\hline
$\nphys=28$\\
0010000001110101010011011010\\
0110110010100110101001010110\\
0111110111011011110101010010\\
0001110000101110010001101111\\
1001011110000111001111000001\\
0000111000011001110100111110\\
\hline
$\nphys=30$\\
010010100111011011011000000110\\
001010110011100100001010111010\\
100001011011011100011100010001\\
011011001001100000110000011001\\
101000010110001101111001100001\\
011010101001001101001111111101\\
\end{tabular}
\caption{Distance $d=4$ codes.  We give $\nstab-1$ stabilizer generators  for each code as bit strings of length $\nphys$; a $1$ in the string in some position indicates that the generator contains the given Majorana operator.  In addition, the fermion parity operator (not shown in the table) is a generator.}
\label{d4gentable}
\end{table}

\section{Table of Codes Found For $d=6$}
\label{d6section}
We give in Table \ref{d6gentable} a table of  the codes found using numerical search for distance $d=6$.

\begin{table}[hbtp]
\begin{tabular}{l}
$\nphys=28$\\
1100110110001100000001000010\\
0001100110010000010100111110\\
0011001010100001100111101101\\
1011100100010101011010111000\\
0000010101110010101010000001\\
0010011100000011101100101001\\
0000101001101111110110111111\\
0010010100101000100101110000\\
1101100001101001101010101001\\
1010010101101110011001110000\\
0101011000001111010111111111\\
\hline
$\nphys=30$\\
011100000111001010010110011100\\
000111111101001001001010011110\\
111001000011101100110011100110\\
100111111010111000111011011101\\
010110101011000011011011110010\\
001100001100111011001110111100\\
100001110110010011100111101010\\
011000000001111110000100011001\\
101001001111000111110011101011\\
000100001001011011111100011001\\
100110000010011101111110100000
\end{tabular}
\caption{Distance $d=6$ codes.  We give $\nstab-1$ stabilizer generators  for each code as bit strings of length $\nphys$; a $1$ in the string in some position indicates that the generator contains the given Majorana operator.  In addition, the fermion parity operator (not shown in the table) is a generator.}
\label{d6gentable}
\end{table}
\end{document}